\documentclass[reprint, amsmath, amssymb, aps, pra, showkeys]{revtex4-2}
\usepackage{csquotes}
\usepackage{graphicx}
\usepackage{dcolumn}
\usepackage{bm}

\usepackage[colorlinks=true, allcolors=blue]{hyperref}

\hypersetup{
    colorlinks=true,
    linkcolor=blue,
    filecolor=magenta,      
    urlcolor=cyan,
    citecolor=blue,
    pdftitle={Nasser et. al., On-Demand},
    pdfpagemode=FullScreen}

\begin{document}

\preprint{APS/123-QED}

\title {Multimode and Random-Access Optical Quantum Memory via Adiabatic Phase Imprinting}

\author{Nasser Gohari Kamel $^{1,2}$}
\author{Sourabh Kumar$^{1,2}$}
\author{Ujjwal Gautam$^{1,2}$}
\author{Erhan Saglamyurek$^{2,3}$}
\altaffiliation{Present address: Lawrence Berkeley National Laboratory and Department of Physics, University of California, Berkeley, California, 94720, USA.}
\author{Vahid Salari$^{1,2}$}
\author{Daniel Oblak$^{1,2}$}%
\email{doblak@ucalgary.ca}
\homepage{https://qcloudlab.com/}

\affiliation{$^{1}$Institute for Quantum Science and Technology, University of Calgary, Calgary, AB, Canada T2N 1N4}
\affiliation{$^{2}$Department of Physics and Astronomy, University of Calgary, Calgary, AB, Canada T2N 1N4}
\affiliation{$^{3}$Department of Physics, University of Alberta, Edmonton, AB, Canada T6G 2E1}

\date{\today}

\begin{abstract}
A photonic quantum memory capable of simultaneously storing multiple qubits and subsequently recalling any randomly selected subset of the qubits, is essential for large-scale quantum networking and computing. Such functionality, akin to classical Random-Access Memory (RAM), has proven difficult to implement due to the absence of a versatile random-access mechanism and limited multimode capacity in existing quantum memory protocols. A potential path to developing the quantum analog to RAM is offered by photon-echo protocols in rare-earth ion-doped materials, such as Revival Of Silenced Echo. These can utilize optical rephasing pulses to selectively read-out frequency multiplexed photonic qubits within an inhomogeneously broadened optical transition. However, the conventional non-adiabatic nature of the rephasing pulses requires intense, short-duration pulses, impeding their fidelity and multimode capacity. To address these critical limitations, we introduce an alternate protocol that employs Rapid Adiabatic Passage (RAP) rephasing pulses, to realize quantum memory, which invokes
phase-imprints to suppress undesirable echoes. Using the optical transitions of a $^{171}{\rm Yb}^{3+}$:${\rm Y}_2{\rm SiO}_5$ crystal, we demonstrate the storage and retrieval of multiple intricate spectro-temporally photonic modes and achieve optical random access memory across eight distinct spectral modes. This protocol yields greatly enhanced mode-mapping versatility while substantially lowering the required rephasing pulse intensity, providing a more efficient and reliable approach for high-fidelity qubit storage and retrieval.
\end{abstract}

\maketitle

\section{Introduction}\label{sec:Main}

The development of efficient quantum memory (QM) systems~\cite{saglamyurek2011broadband, de2008solid, duranti2024efficient}, enabled by precisely engineered light-matter interactions, underpins the feasibility of the Quantum Internet~\cite{kimble2008quantum}, long-distance quantum communication~\cite{duan2001long}, and distributed quantum computing~\cite{Knaut2024573,liu2023quantum, de2019massively}, which is fundamental to advancing quantum information science~\cite{ladd2010quantum,  zhang2024realization, ekert1991quantum, lago2023long}. A practical QM must offer on-demand and, ideally, random access recall of multiple simultaneously stored photonic qubit modes. Such random-access QM is critical for quantum computing architectures~\cite{giovannetti2008quantum, park2019circuit, gouzien2021factoring} and for achieving high communication rates in quantum repeaters over long distances~\cite{sangouard2011quantum}. A range of QM protocols have been explored~\cite{kraus2006quantum, Saglamyurek2018774,julsgaard2004experimental,guo2019high}, including Electromagnetically Induced Transparency (EIT)~\cite{ma2017optical} and Atomic Frequency Comb (AFC)~\cite{afzelius2009multimode}, each presenting distinct advantages and challenges in quantum state storage and retrieval~\cite{lvovsky2009optical, heshami2016quantum, tittel2010photon, rastogi2019discerning}. With AFC-QM, a type of echo-based protocol, storage and retrieval of large number of optical qubits has been demonstrated~\cite{businger2022non, wei2024quantum}, while on-demand storage utilizing strong control fields (spin-wave storage) is limited to a few modes~\cite{Laplane2016, Ortu2022,Cho2016100}. The restricted multimode capacity is largely due to finite bandwidth and the Rabi frequency of the optical control fields. Moreover, all the stored modes are simultaneously retrieved, thus, not realizing a random-access QM. The Hahn Echo memory, or two-pulse photon-echo (2PPE)~\cite{hahn1950spin}, offers a more straightforward implementation of a memory protocol, however, it is unsuitable for QM implementation due to signal amplification and spontaneous emission from the inverted medium~\cite{ruggiero2009two}. The Revival Of Silenced Echo (ROSE) protocol addresses this by suppressing the initial/primary amplified echo and invoking a secondary echo from an uninverted transition, demonstrated using a pair of spatially mismatched, strong optical control fields to store an optical input mode~\cite{damon2011revival}. 

Here we introduce a QM protocol that employs two identical Rapid Adiabatic Passage (RAP) pulses to achieve re-emission of stored excitations~\cite{pascual2013securing}. The suppression of the primary echo is provided by the phase-imprint left by the first RAP pulse, which subsequently is canceled out by the second co-propagating RAP pulse, allowing for the emission of the secondary echo only. Key advantages of using RAP pulses are their reduced power requirement and better resilience to frequency and power fluctuations compared to conventional short~$\pi$-pulses~\cite{demeter2013adiabatic, lauro2011adiabatic, garwood2001return, loy1974observation}. This technique has proven valuable in nuclear spin systems~\cite{dumez2021frequency, demeter2013adiabatic}, where random-access QM has been demonstrated across five storage modes utilizing microwave photons~\cite{o2022random}. To our knowledge, this work is the first demonstration of RAP pulses meeting adiabaticity conditions~\cite{pascual2013securing} to successfully revive optical silenced echoes. We conduct an in-depth experimental and theoretical study of on-demand re-emission, memory efficiency, and multimode excitations using RAP pulses in the optical domain. Our investigation explores multiplexed encoding schemes across temporal, spectral, and simultaneous spectro-temporal modes to demonstrate different aspects of the memory protocol. We demonstrate that multitone RAP pulses facilitate bespoke storage and on-demand retrieval of multiple spectro-temporal modes to meet a core random-access QM requirement and enable random-access operations. While optical random-access QM has previously been achieved via EIT with spatial modes~\cite{zhang2024realization, messner2023multiplexed}, our Rapid Adiabatic Passage Phase Imprint (RAPPI) memory protocol distinctively draws on more scalable multiplexed encoding in the low-photon-number regime, establishing a robust framework for QMs.

\begin{figure*}[htbp]
  \centering
\includegraphics[width=1\textwidth]{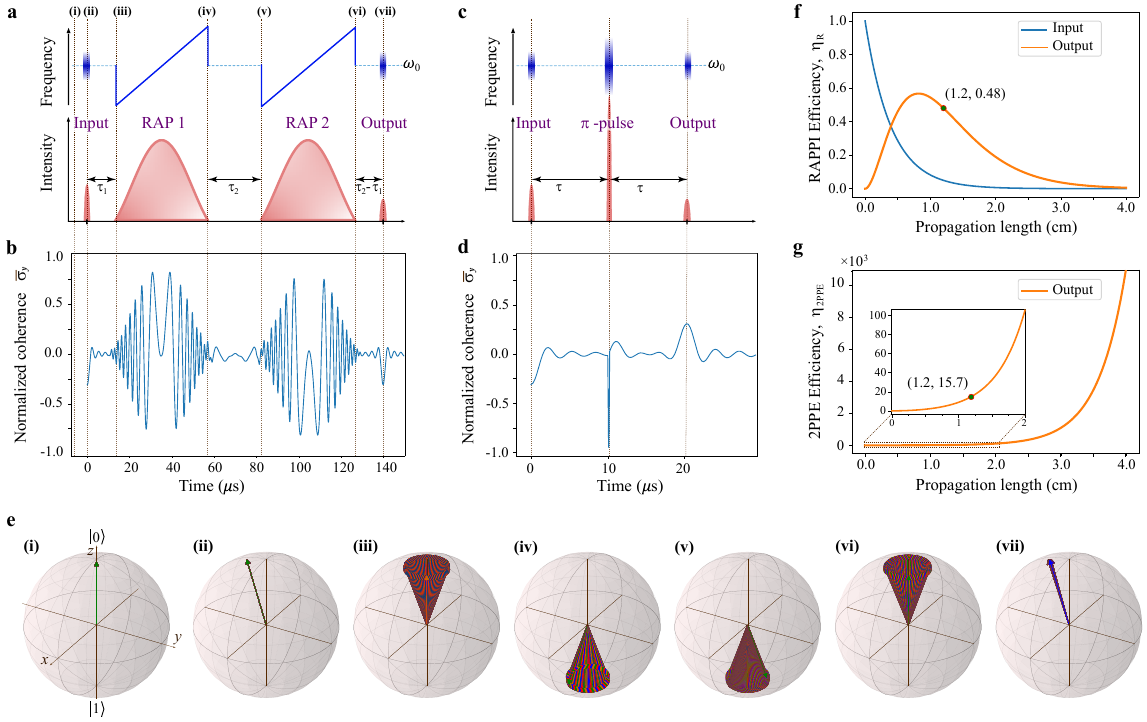}
\caption{Comparison of RAPPI and 2PPE protocols. \textbf{a,} RAPPI protocol pulse-sequence showing the frequency span (top) and intensities (bottom) of the probe and the RAP pulses. Two identical RAP pulses (RAP1 and RAP2) are applied sequentially, with the first one imprinting distinct detuning-dependent phase onto the atoms to suppress the primary echo and the second RAP pulse rephasing the atoms to enable on-demand re-emission of the stored probe-pulse. \textbf{b,} Normalized coherence evolution (\( \bar\sigma_y \)), showing primary echo suppression after RAP1 and emission of secondary echo post-RAP2. \textbf{c,} 2PPE pulse sequence showing the frequency span (top) and intensity (bottom) of the probe, strong \( \pi \)-pulse, and primary echo, with their frequency centered around \( \omega_0 \). \textbf{d,} Normalized coherence evolution (\( \bar\sigma_y \)) in the 2PPE protocol where a strong \( \pi \)-pulse rephases the coherence after \( \tau \) to emit the primary echo. \textbf{e,} Evolution of Bloch vectors in RAPPI Protocol: \textbf{(i)} Initially, all the atoms are in the ground state. \textbf{(ii)} After the absorption of the probe-pulse, the Bloch vectors rotate by a small angle about the X-axis, and \textbf{(iii)} undergo free evolution up to time \( \tau_1 \). \textbf{(iv)} RAP1 inverts the population while imprinting a detuning-dependent phase to each vector. \textbf{(v)} Free evolution continues during the interval \( \tau_2 \). \textbf{(vi)} RAP2 brings the vectors near the ground state while adding an opposite detuning-dependent phase than RAP1. \textbf{(vii)} The Bloch vectors then evolve over \( \tau_2 - \tau_1 \), culminating in full rephasing. \textbf{f, g,} Simulated memory efficiencies versus propagation length for the RAPPI and 2PPE protocols, respectively. The inset in \textbf{g} illustrates the efficiency of the 2PPE protocol for a short propagation length, showing that, for a given propagation length, 2PPE exhibits amplification relative to the RAPPI protocol.}
\label{fig1:MemoryProtocol}
\end{figure*}

\section{Theoretical description}\label{sec:Theoretical_description}
\subsection{Protocol at a glance}

The RAPPI memory protocol, schematically depicted in Fig.~\ref{fig1:MemoryProtocol}\textbf{a}, begins with the probe-pulse, to be stored, entering the inhomogeneously broadened medium at \( t = 0 \). After a delay \( \tau_1 \), the first RAP pulse (RAP1) characterized by a duration \( \tau_R \) and a sweep rate \( R \) around a center frequency of $\omega_0$, is applied. Following a longer delay \( \tau_2 > \tau_1 \), a second identical RAP pulse (RAP2) is introduced. The probe-pulse, recalled as a secondary echo, is subsequently detected at a time \( \tau_2 - \tau_1 \) following the RAP2 pulse yielding a total storage time of $t_E = 2(\tau_2+\tau_R)$. Importantly, no primary echo is permitted between the RAP pulses due to the uncompensated phase-imprint from the RAP1 pulse.

In the following theoretical description of the protocol, first, we model the evolution of the atomic ensemble given a RAP pulse on resonance with the inhomogeneously broadened atomic transition. Second, we describe the absorption and subsequent re-emission of the weak optical probe-pulse by the same atomic ensemble. A more detailed theoretical analysis is provided in Supplemental Material~\cite{Nasser2025Multimode} with complementary material in~\cite{Minar2010,pascual2013securing, o2022random}.

\subsection{RAP Hamiltonian}\label{sec:RAPHam}
The evolution of an atom in the ensemble during the application of the RAP pulses is modeled by a simplified Hamiltonian~\cite{o2022random}, represented in a frame rotating at the center frequency of the RAP pulse, \( \omega_0 \):
\begin{equation}
\label{eq:Hamil}
    \hat{H}_R = -\frac{\delta}{2}\hat{\sigma}_z + \frac{\Omega_{R} (t)}{2} \left( e^{-i \phi (t)} \hat{\sigma}^+ + e^{i \phi (t)} \hat{\sigma}^- \right).
\end{equation} 

Each atom within the inhomogeneous linewidth has a unique transition frequency, \( \omega_a \), with detuning $\delta = \omega_a-\omega_0$. Thus, the Hamiltonian’s first term describes the atomic energy detuning (relative energy of an individual atom).
The second term represents the atom-light interaction. Here, \( \Omega_R(t) \) is the time-dependent Rabi frequency of the RAP-pulse, which in turn is determined by its temporal profile.
This is characterized by a chirp rate \( R \), with its phase evolving over time as \( \phi(t) = \frac{1}{2} R(t - t_0)^2 \), where \( t_0 \) denotes the pulse’s temporal center. The operators \( \hat{\sigma}^+ \) and \( \hat{\sigma}^- \) correspond to atomic excitation and de-excitation, facilitating transitions between the ground and excited states. 

A similar approach is used to model the ensemble evolution under the application of a pair of short and intense monochromatic $\pi$-pulses as employed for 2PPE, illustrated in Fig.~\ref{fig1:MemoryProtocol}\textbf{c}.

\begin{figure*}[htbp]
\centering
\includegraphics[width=0.9\textwidth]{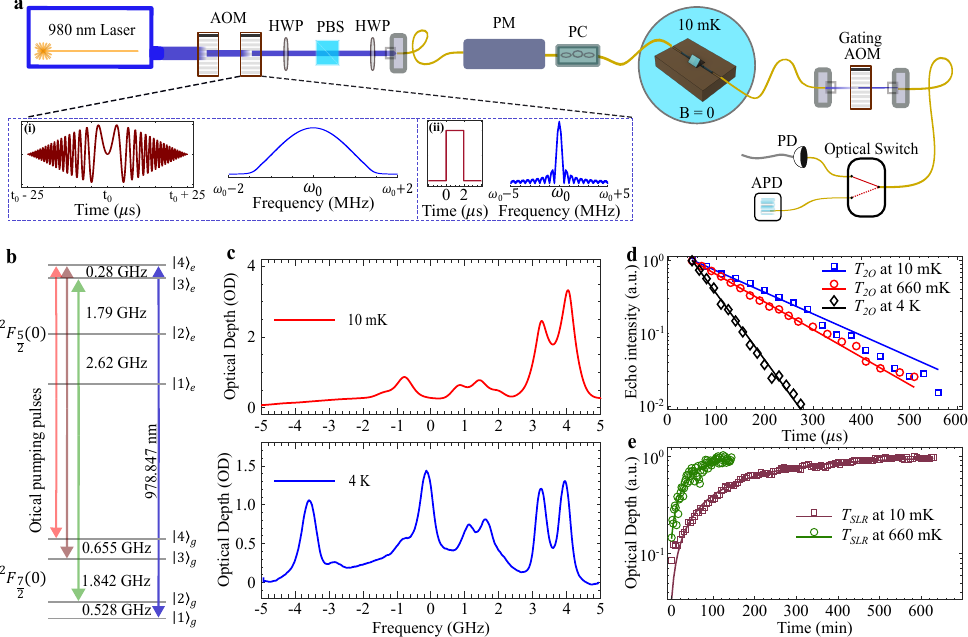}
\caption{Experimental setup and memory medium characteristics. \textbf{a,} HWP: Half wave plate, PBS: Polarizing beam splitter, PM: Phase Modulator, PC: Polarization Controller, APD: Avalanche Photodetector, and PD: Photodetector. Two AOMs are placed in series to generate probe and RAP pulses from a CW 980~nm laser source. PM is used to shift the laser frequency for optical pumping. A 12~mm long Yb:YSO crystal is mounted on a custom alignment-free mount and cooled to 10~mK with light coupled in and out with optical fibers and gradient-index (GRIN) lenses. To ensure most efficient light-matter interaction, the polarization of light is fixed along ${D_2}$ crystallographic axis~\cite{tiranov2018spectroscopic}. The third AOM gates the detectors from intense RAP pulses. Probe and re-emitted pulses are detected by a high-gain 10~MHz bandwidth PD for classical pulses or an APD for single-photon regime pulses. Inset \textbf{(i)} and \textbf{(ii)} are amplitudes in time and frequency domains for the RAP and the probe-pulses respectively. Frequency of the RAP pulse is linearly swept with $2\pi \times 30$~MHz/ms over its center frequency $\omega_{0}$ where it perfectly covers the spectral bandwidth of a 2~$\mu$s input probe-pulse with the same center frequency. \textbf{b,} Yb:YSO energy levels for site II ions at zero magnetic field. $|1_g\rangle  \rightarrow |4_e\rangle $ (blue arrow) is the storage transition. $|2_g\rangle  \rightarrow |3_e\rangle $ (green), $|3_g\rangle  \rightarrow |4_e\rangle $ (brown), and $|4_g\rangle  \rightarrow |4_e\rangle $ (red) transitions are used for optical pumping process. \textbf{c,} Absorption spectra of Yb:YSO at two different temperatures. At 10 mK, population of thermal phonons in the thermal bath is low, leading to natural spin polarization in the lowest two ground states. \textbf{d,} Optical coherence time results measured by 2PPE method at 10 mK (585.9~$\mu$s), 660~mK (457.5~$\mu$s), and 4~K (190.41~$\mu$s) using $I(t)=I_{0}\ \text{exp}(-4t/T_{2O})$ to fit the experimental data. \textbf{e,} To extract the SLR time at 10 mK (210 min) and 660 mK (55 min) the population of $|1_g\rangle$ is optically pumped to other ground states and a time resolved optical depth of $|1_g\rangle$ is monitored. The results are fitted to a single exponential functions as $OD(t)=1-a\ \text{exp}(-t/T_{SLR})$ to extract $T_{SLR}$.}
\label{fig2:Setup}
\end{figure*}
\subsection{Probe-pulse absorption and recall}\label{sec:probeevolve}

The efficiency of probe re-emission can be determined by solving the Maxwell-Bloch equations~\cite{crisp1970propagation} for the state of the atomic ensemble and the probe-pulse. A weak input probe signal will decrease exponentially with propagation distance \(z\), such that its Rabi frequency \(\mathcal{S}\) varies with distance according to~\cite{ruggiero2009two, crisp1970propagation}:
\begin{equation}
\label{eq:inp_field}
    \partial_z \mathcal{S}(z,t) = -\frac{\alpha}{2 \pi} \int g (\omega_{a}) \frac{\langle \hat\sigma_z (\omega_{a};z,t=0)\rangle}{2} \ {\mathcal{S} (z,t)}\ d\omega_{a} ,
\end{equation}
where $\langle\hat\sigma_z\rangle$ identifies the atomic population in the ground state, \(g(\omega_{a})\) is the spectral profile of the inhomogeneous broadening, and \(\alpha\) corresponds to the peak optical depth of the crystal in the light propagation direction. If all atoms are initially in the ground state, \( \langle\hat\sigma_z (\omega_{a};z,0)\rangle = 1\), and the absorption profile is uniform, then the above equation simplifies to $\partial_z \mathcal{S}(z,t)= -\frac{\alpha}{2} {\mathcal{S}(z,t)}$. 

As the probe-pulse traverses the medium, it generates coherence within the atoms, quantified by the expectation value of the $\hat \sigma_y$ operator. Both $\langle\hat\sigma_y\rangle$ and $\langle\hat\sigma_z\rangle$ evolve during the application of the RAP pulses (as governed by the Hamiltonian in Eq.~\ref{eq:Hamil}) as well as during the wait times. At the time of echo emission, coherence is converted back to the optical field according to:

\begin{equation}
\label{eq:out_field}
\begin{split}
    \partial_z \mathcal{E}(z,t) =
    -\frac{\alpha}{2 \pi} \int g (\omega_{a}) \bigg(\frac{\langle \hat\sigma_z (\omega_{a};z,t_E)\rangle}{2} {\mathcal{E} (z,t)} \\
    - \langle \hat\sigma_y (\omega_{a};z,t_E)\rangle\bigg) \ d\omega_{a}
\end{split}
\end{equation}

Here, \(\mathcal{E}\) denotes the Rabi frequency of the output field, \(t_E\) is the echo time, and $\langle \hat\sigma_y (\omega_{a};z,t_E)\rangle$ signifies the coherence at the echo time.\par

\subsection{Memory protocol mapping on Bloch sphere}
To illustrate the protocol, we map the evolution of the state of our atomic ensemble -- from probe-pulse absorption over RAP pulses to the probe echo (Fig.~\ref{fig1:MemoryProtocol}\textbf{a})  -- on the Bloch sphere as shown in Fig.~\ref{fig1:MemoryProtocol}\textbf{e}. Each atom in the ensemble is indexed by $j \in \{1,...,N\}$, and is represented by a two-dimensional complex state-vector $|\psi_i^{(j)}\rangle$ in the Hilbert space spanned by the ground and excited state. The inhomogeneous broadening of the transition energies is accounted for by assigning different detunings $\delta_j = \omega_a^{(j)}-\omega_0$ to each of the $N$ Bloch vectors used to model the ensemble state. This also allows us to continuously track the ensemble average coherence given by $\overline{\sigma}_y = \frac{1}{N}\sum_{j=1}^N \langle \hat\sigma_y^{(j)} \rangle$, where $\langle \hat\sigma_y^{(j)} \rangle \equiv \langle \psi^{(j)}|\hat\sigma_y |\psi^{(j)}\rangle$, as plotted in Fig.~\ref{fig1:MemoryProtocol}\textbf{b}. 
For comparison, we also simulate $\overline{\sigma}_y$ for the 2PPE memory protocol (Fig.~\ref{fig1:MemoryProtocol}\textbf{c}) and plot its time evolution in Fig.~\ref{fig1:MemoryProtocol}\textbf{d}.
\par

To visualize the RAPPI QM protocol using Bloch sphere representation, the initial state of the ensemble (choosing $N=1001$) with all atoms in the ground state is depicted in Fig.~\ref{fig1:MemoryProtocol}\textbf{e}-\textbf{i}. The probe-pulse induces a small excitation, causing identical small rotations of all the $N$ Bloch vectors around the X-axis (Fig.~\ref{fig1:MemoryProtocol}\textbf{e}-\textbf{ii}). Next, due to inhomogeneous broadening, different atoms acquire a time-dependent phase during the waiting time \( \tau_1 \) causing the $N$ Bloch-vectors to mutually dephase (Fig.~\ref{fig1:MemoryProtocol}\textbf{e}-\textbf{iii}). RAP1 is then applied (Fig.~\ref{fig1:MemoryProtocol}\textbf{e}-\textbf{iv}), inducing an adiabatic \(\mathrm{\pi}\) rotation about the X-axis (population inversion) and imprinting a frequency-dependent phase to the Bloch vector of each atom~\cite{Conolly199128, pascual2013securing}. In essence, the frequency chirp of the RAP pulse drives atoms with different inhomogeneous-detunings at slightly offset times leading to a distinct phase of each atom. This phase imprint distinguishes the RAPPI protocol from the 2PPE method in which the population transfer to the excited state lacks the phase-imprint and, thus, the atomic state rephases at time \( \tau_1 \) following the first \(\mathrm{\pi}\)-pulse (while the medium is inverted), leading to an echo of the input probe-pulse as manifested by the coherence peak in Fig.~\ref{fig1:MemoryProtocol}\textbf{d}. In the RAPPI protocol, the additional phase from the RAP pulse prevents this rephasing, resulting in the silencing of the primary echo as evident from the lack of a coherence peak in Fig.~\ref{fig1:MemoryProtocol}\textbf{b}.\par

After a wait-time of \( \tau_2 \) (FIG.~\ref{fig1:MemoryProtocol}\textbf{e}-\textbf{v}), the application of RAP2 (identical to RAP1) adiabatically transfers the atomic states (Bloch vectors) back towards the ground state (Fig.~\ref{fig1:MemoryProtocol}\textbf{e}-\textbf{vi}), while again imprinting a detuning-dependent phase on each atom. This inherently cancels the phase-imprint of RAP1. 
Hence, at time \( \tau_2 - \tau_1 \) following RAP2, all atoms undergo collective rephasing of their Bloch-vectors (Fig.~\ref{fig1:MemoryProtocol}\textbf{e}-\textbf{vii}), effectively completing a \(2\mathrm{\pi}\) rotation about the X-axis from their initial state (Fig.~\ref{fig1:MemoryProtocol}\textbf{e}-\textbf{ii}), which results in a strong echo emission of the probe-pulse. This is also reflected in the coherence as a peak matching that of the input pulse (Fig.~\ref{fig1:MemoryProtocol}\textbf{b}).

Another distinction between RAPPI and 2PPE is depicted in Fig.~\ref{fig1:MemoryProtocol}\textbf{f}, \textbf{g}, wherein the RAPPI protocol reaches a peak efficiency of $\sim$~54\% (near the theoretical maximum for two-level absorptive QMs without cavity enhancement), while the 2PPE protocol exhibits amplification, with efficiency exceeding 100\%.

\begin{figure*}[htbp]
  \centering
    \includegraphics[width=1\linewidth]{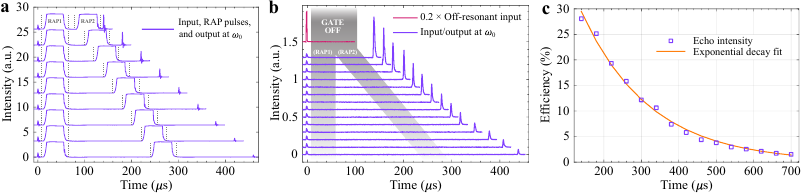}
\caption{RAPPI protocol and efficiency. \textbf{a,} Experimental results depicting storage and re-emission of a 2~$\mu$s rectangular probe-pulse with a peak power of 1.75~$\mu$W together with a pair of intense RAP pulses saturating the PD. The vertical grey lines around the RAP pulses mark the gate off times for further experiments. Here, $\tau_{1}$ is fixed at 10~$\mu$s while $\tau_{2}$ is varied from 20-300~$\mu$s, resulting in an effective delay of 140-700~$\mu$s for re-emission from the input probe-pulse (results shown up to 480~$\mu$s). No echo between the identical RAP pulses illustrates the silencing of echo due to RAP1 and echo after RAP2 demonstrates rephasing and re-emission of the stored probe-pulse. \textbf{b,} Demonstration of RAPPI protocol with gating the RAP pulses. Storage and recall of a probe-pulse (peak power of 350 nW) with off-resonant probe-pulse are recorded to evaluate the storage efficiency. Echo emission occurs at time $\tau_{2} - \tau_{1}$ post-RAP2. \textbf{c,} Storage efficiency as a function of storage time. Efficiencies are quantified as the ratio of echo intensity to off-resonant probe-pulse intensity. The results are fitted to $\eta = \eta_\mathrm{R} \exp\left({-2t}/{T_{2M}}\right)$, yielding a memory coherence time of $T_{2M}=365$~$\mu$s.}
  \label{fig:efficiency}
\end{figure*}

\begin{figure*}[htbp]
    \centering
    \includegraphics[width=\linewidth]{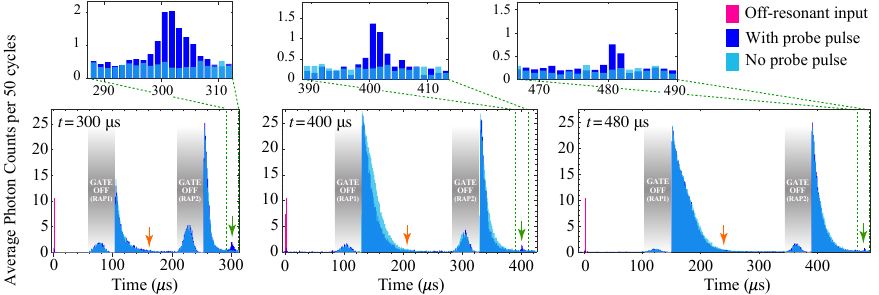}
    \caption{Single-photon level probe-pulse storage. Average photon count histograms over 50 cycles, comparing storage with (blue) and without (sky blue) a weak coherent probe-pulse for three different storage times. The weak coherent input pulse at off-resonant is depicted in hot-pink color for each histogram at $t=0$. Despite gating the RAP pulses, some leak light is detected at their position. The detection Gate remains off throughout RAP pulses and is activated only 3~$\mu$s after application of these pulses. In each histogram, green arrows indicate secondary echo while orange arrows are used to demonstrate the position of primary silenced echo. Insets: Zoom in view of the echo window showing noise and echo photon counts. The resulting storage efficiencies are 13\%, 9\%, and 5\% for 300~$\mu$s, 400~$\mu$s, and 480~$\mu$s storage times, respectively. These values are comparable to the efficiencies achieved for storage using a classical probe-pulse and are calculated as $\eta(\%) = \left[(\text{echo intensity} - \text{noise level}) / (\text{intensity of the off-resonant input})\right] \times 100$.}
     \label{fig:Single_photon_level}
\end{figure*}

\section{Experimental setup and memory medium}\label{sec:Experimental_Setup_and_Memory_Crystal}

We experimentally implement the RAPPI protocol utilizing a 5~ppm isotopically enriched $^{171}{\rm Yb}^{3+}$ doped ${\rm Y}_2{\rm SiO}_5$ (Yb:YSO) crystal held in a cryostat. The experimental setup, illustrated in FIG.~\ref{fig2:Setup}\textbf{a}, is configured for single-pass optical transmission through the crystal. All optical signals, including classical and weak coherent probe and RAP pulses, are derived from a continuous-wave 980~nm laser, combined with two acousto-optic modulators (AOMs) in series for tailoring amplitudes and spectral profiles (inset in FIG.~\ref{fig2:Setup}\textbf{a}).\par

The $^{171}{\rm Yb}^{3+}$ ions, possessing electronic and nuclear spins of $1/2$, exhibit hyperfine splitting that results in four distinct ground and excited states. The optical transitions are centered at 978.854~nm, as shown in Fig.~\ref{fig2:Setup}\textbf{b}. Absorption spectra of our sample is measured at two temperatures in Fig.~\ref{fig2:Setup}\textbf{c}. At 10~mK, phonon population with energies compatible to hyperfine splitting are negligible~\cite{welinski2020coherence} enabling spin polarization within the lowest energy states ($|1_g\rangle$ and $|2_g\rangle$) where $|1_g\rangle \leftrightarrow |4_e\rangle$ transition is employed to execute the storage protocol. While prior studies have characterized Yb:YSO properties at higher temperatures (2–3~K)~\cite{tiranov2018spectroscopic, businger2022non}, our systematic investigation from 10~mK to 4~K, reveals a threefold increase in optical coherence ($T_{2O}$, $\sim 0.6$~ms~\cite{chiossi2024optical}) and spin-lattice relaxation (SLR) time of over 3 hours at 10~mK, as shown in Fig.~\ref{fig2:Setup}\textbf{d} and~\ref{fig2:Setup}\textbf{e}. This underlines the suitability of Yb:YSO as a quantum light-matter interface. Notwithstanding the long SLR, we apply optical pumping to initialize the spin population in the $|1_g\rangle$ state prior to all storage experiments, providing a controlled baseline for all measurements. 

To implement the RAPPI protocol, we synthesize the RAP pulses, depicted in FIG.~\ref{fig2:Setup}\textbf{a}-\textbf{i}, using a novel dynamic combination of time-dependent modified Sinc amplitude and linear phase modulation functions as follows:
\begin{equation}
    P(t)= A(t) \sin (\omega_0 t + \phi (t)),
    \label{eq:0}
\end{equation}
with the phase and amplitude terms given by
\begin{align}
    A(t)&= A_0 \text{sinc}\left(\frac{2(t-t_0)}{\tau_R} \right), \\  
    \phi (t)&= \frac{\Delta_{R}}{\tau_R} (t-t_0)^2 \ ,
    \label{eq:1}
\end{align}
where $A_0$ represents the peak pulse amplitude, $\tau_R$ the total pulse duration, and $\Delta_{R}$ defines the frequency chirp range, such that the chirp rate is given by $R=\Delta_{R}/\tau_R$.\par

RAP pulses enable precise manipulation of quantum states in electronic and nuclear spin systems through optimized pulse shape and frequency sweep, allowing efficient population transfer with minimal unwanted transitions~\cite{demeter2013adiabatic, lauro2011adiabatic, garwood2001return, loy1974observation}. These pulses are optimized with respect to several experimental conditions. First, the frequency span of the RAP pulse should closely match the frequency spectrum of the input pulse (FIG.~\ref{fig2:Setup}\textbf{a}-\textbf{i} and \textbf{ii}). This necessitates that the relationship between the frequency bandwidths of the input pulse ($\Delta\nu_{\text{input}}$) and RAP pulses ($\Delta\nu_{\text{RAP}}$) satisfy the condition \( \Delta\nu_{\text{RAP}} \gtrsim 3\Delta\nu_{\text{input}} \). Additionally, to ensure optimal conditions for efficient operation of the RAPPI protocol, the adiabaticity condition \( \Omega_R^2 \gg R \) must be met, where \( \Omega_R \) represents the resonant Rabi frequency of the RAP-pulse. The RAP pulse optimization is elaborated in Supplemental Material~\cite{Nasser2025Multimode}.

A key distinction between the RAPPI protocol implementation in the microwave and optical domains stems from the significantly different inhomogeneous broadenings of the respective nuclear-spin and electronic transitions. The narrow inhomogeneous broadening of the nuclear spins~\cite{pascual2013securing}, can easily be matched with the bandwidth of an input microwave probe-pulse. The RAP pulse, thus, interacts with the entire ensemble storing the probe-pulse coherence. Conversely, our optical probe-pulses of a few~$\mu$s duration only cover a fraction of the extensive inhomogeneous broadening of the optical transitions in rare-earth ion-doped crystals. Hence, the RAP pulse, swept over $\Delta_R$, will interact with \enquote{spectator atoms} that have not interacted with the probe-pulse. Nonetheless, if ion-ion interactions are negligible, we find that the RAPPI protocol can still be effectively implemented in the optical domain. Further, experimental characterization of the memory crystal and theoretical simulations regarding the effects of spectator atoms are provided in the Supplemental Material~\cite{Nasser2025Multimode}.\par

\section{Experimental realization of memory protocol}\label{sec:Experimental_realization_of_memory_protocol}

To demonstrate the RAPPI protocol and characterize its efficiency, a weak, short probe-pulse (peak powers ranging from 0.35 to 1.75~$\mu$W) is absorbed by our sample followed by two identical RAP pulses, each with a duration of 50~$\mu$s, peak power of 400~$\mu$W, Rabi frequency of $\Omega_R=2\pi\ \times 0.35$~MHz, and sweep rate of $R=2\pi \times 30$~MHz/ms. FIG.~\ref{fig:efficiency}\textbf{a} presents the recorded experimental results for the storage and recall of an input probe-pulse from $t_E=140~\mu s$ to $460~\mu s$, while storage duration can reach up to $700~\mu s$. In addition to a small amount of the unabsorbed probe-pulse, visible at $t=0$, and the RAP pulses that saturate the detector, only the recalled pulse is visible. Notably, no recall occurs after RAP1, thus, confirming the echo-suppression due to the phase-imprint. FIG.~\ref{fig:efficiency}\textbf{b}, shows the results using the same RAP pulse parameters, with another AOM at the output of the memory gating the RAP pulses to avoid detector saturation (all subsequent measurements employs such gating). Normalizing the recalled pulse to the off-resonantly transmitted probe-pulse, we deduce an efficiency of 28\% at a storage time of $t=140~\mu$s. The efficiency of the RAPPI protocol as a function of storage time exhibits single exponential decay, yielding a memory decay time of \( 365.09~\mu\text{s} \) (FIG.~\ref{fig:efficiency}\textbf{c}).\par

Predicted efficiency from our model of the RAPPI protocol with $z=1.2$ cm long medium -- the length of our Yb:YSO crystal -- is $\eta_\mathrm{R}=48\%$ (as indicated in FIG.~\ref{fig1:MemoryProtocol}\textbf{f}). However, our model, so far, has not incorporated dephasing due to the finite coherence time, $T_{2O}$, of the optically excited level. Including the effect of the finite coherence time, measured to be 586~$\mathrm{\mu}$s (Fig.~\ref{fig:efficiency}d), we estimate the theoretical efficiency of re-emission as $\eta_\mathrm{R}= e^{-2 t_E/T_{2O}}$ = 29.7$\%$, which is in excellent agreement with the experimental efficiency of 28$\%$.

For the RAPPI protocol to function as a quantum memory, it is essential to demonstrate the storage and retrieval of probe-pulse at the single-photon level. FIG.~\ref{fig:Single_photon_level} depicts the experimental realization of the protocol for storage and recall of a weak coherent probe-pulse containing about 2500 photons per pulse at the memory input. This is estimated from the detection of the transmitted probe-pulse, off-resonant with the atomic ensemble, and accounting for the known losses of optical elements up until the single-photon detectors (see Methods and Supplemental Material~\cite{Nasser2025Multimode}). We achieve an efficiency of up to 13\% for 300~$\mu$s storage duration and are able to detect a secondary echo up to 480~$\mu$s storage.

Our current experimental configuration, in which the RAP pulses propagate co-linearly with the probe-pulses, prevents lower photon-number storage due to the exponentially decaying Free Induction Decay (FID) associated with RAP pulses. The FID is clearly visible in FIG.~\ref{fig:Single_photon_level} even in the absence of an input probe-pulse (light-blue histograms). It is also evident that the effect of FID reduces the later the secondary echo emission is from the RAP2 pulse, i.e., the larger $\tau_2-\tau_1$ is. However, the signal-to-noise ratio (SNR) is not improved due to the coinciding decay of the emitted probe-pulse echo. The SNR due to FID is elaborated in our detailed theoretical analysis of the RAPPI protocol in the Supplemental Material~\cite{Nasser2025Multimode}. From an experimental angle, the FID can be greatly minimized by adding a small offset between spatial modes of the probe-pulse and RAP pulses as common in on-demand echo-protocols~\cite{bonarota-2014}.

\subsection{Multimode and random access optical memory}\label{Multimode_capacity_of_RAPPI_protocol}

\begin{figure*}[htbp]
   \includegraphics[width=\linewidth]{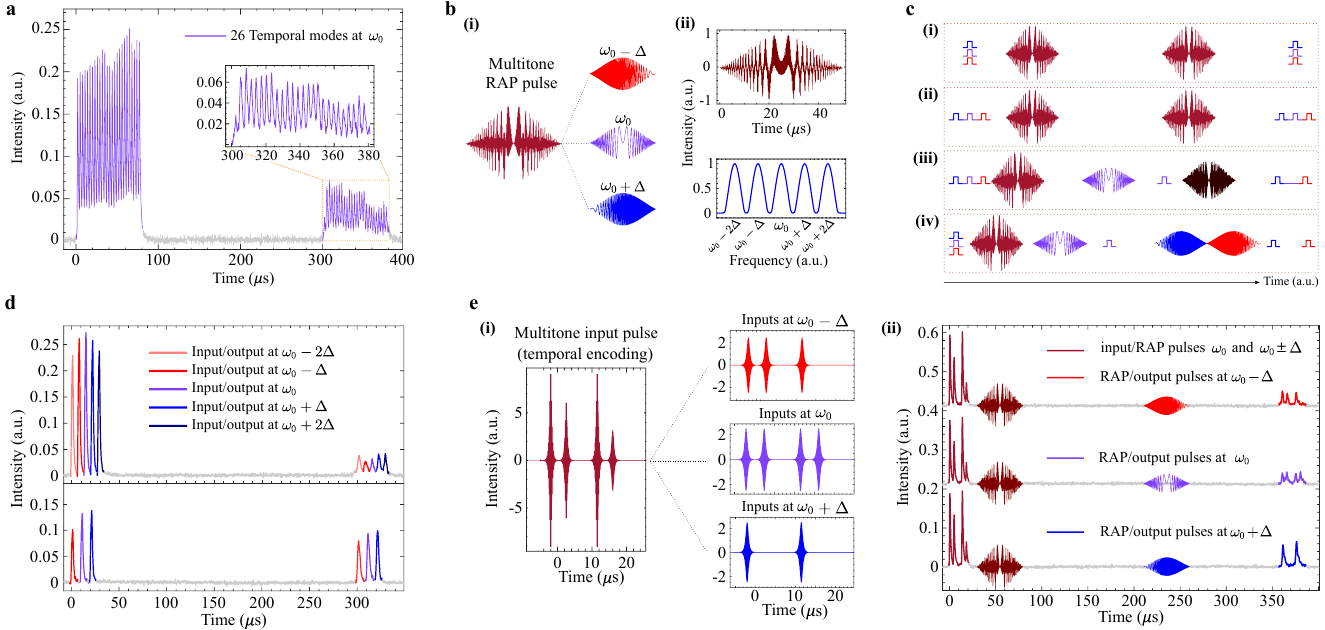}
    \caption{Multimode capacity of RAPPI protocol. \textbf{a,} Storage and retrieval of 26 temporal modes of pulse width 1~$\mu$s and 2~$\mu$s separation. The central frequency of all the input temporal modes and RAP pulses is $\omega_{0}$. RAP pulses are frequency-swept by 4~MHz across $\omega_{0}$ at a rate of $2\pi \times 80$~MHz/ms. \textbf{b,} Multitone RAP pulses in time and spectral domains. \textbf{(i)} Representation of three-tone and \textbf{(ii)} five-tone RAP pulses with each tone's central frequency separated by $\Delta = 2\pi \times 3.5$~MHz. Each constituent tone targets distinct frequency bins (memory cells) for spectrally multiplexed storage and retrieval. RAP pulse for each tone is swept for 1.5~MHz to cover the bandwidth of the individual input tones while not overlapping with other tones. \textbf{c,} Schematic of four key scenarios of the RAPPI protocol for spectro-temporal multimode storage. The first three-tone RAP pulse (common to all scenarios) imprints phase across all the modes, while on-demand and selective retrieval of different tones can be performed with tailored RAP pulses. \textbf{(i)} A three-tone probe-pulse is stored and subsequently retrieved, with all frequency components recalled simultaneously using the same three-tone RAP pulse. \textbf{(ii)} Three temporal modes with distinct tones are stored and retrieved by three-tone RAP pulses in First-In-First-Out (FIFO) scheme. \textbf{(iii)} Three temporal modes with distinct tones are stored. The central spectral mode is retrieved at first (selective and on-demand) by applying the corresponding RAP pulse. The other two modes are retrieved by the corresponding two-tone RAP pulse in FIFO scheme. \textbf{(iv)} A three-tone probe-pulse is stored, yet retrieved selectively and on-demand by applying the corresponding RAP pulses. \textbf{d,} Experimental results of FIFO storage and retrieval of five-distinct tones (top) and three-distinct tones (bottom) based on \textbf{c (ii)} scheme. \textbf{e,} Experimental results of storage and retrieval of simultaneous spectro-temporal multiplexed mode. \textbf{(i)} Preparation of random temporal bins across three spectral modes ($\omega_{0}$ and $\omega_{0}\pm\Delta$) and resulting spectro-temporal multiplexed input probe-pulse. \textbf{(ii)} Experimental traces showing retrieval of selective frequency modes using corresponding RAP pulses. Representation of RAP pulses are depicted to illustrate
    their temporal location.}
  \label{fig:temporalmodes}
\end{figure*}

A highly multimode QM is a fundamental component in quantum repeaters~\cite{simon2007quantum}. Leveraging the $\Gamma_{\text{inh}} \approx 550$~MHz inhomogeneous broadening of Yb:YSO~\cite{tiranov2018spectroscopic}, the RAPPI protocol is in principle able to store photonic probe-pulses, $\tau_{\text{in}}=2.5/\Gamma_{\text{inh}} = 5$~ns duration~\cite{ortu2022multimode}, therefore, increasing temporal multimode capacity. However, limited laser power in our experiment prevents us from driving such high Rabi frequency that is required to address this inhomogeneous broadening. Furthermore, instantaneous spectral diffusion and other decoherence mechanisms associated with increased memory bandwidth or multimode capacity, have been observed to degrade the efficiency of on-demand QMs~\cite{dajczgewand2015optical}.\par 

To evaluate the RAPPI protocol's potential for storage of multimode photonic-excitations, we examine temporal, spectral, and simultaneous spectro-temporal multiplexing strategies. First, for temporal multiplexing we achieve the storage and recall of a train of 26 distinct modes (FIG.~\ref{fig:temporalmodes}\textbf{a}). Detector bandwidth limitations necessitate us to prolong the probe-pulses' duration and spacing, thus, restricting the full multimode capacity. To further expand the memory's multimode capacity, we introduce multitone RAP pulses (FIG.~\ref{fig:temporalmodes}\textbf{b}). These allow different spectral ranges within the atomic inhomogeneous broadening to serve as individual memory ensembles/cells, enabling spectro-temporal storage and retrieval of multiplexed photonic states. Importantly, multitone RAP pulses not only enhance storage capacity but also provide on-demand access to individual memory cells. FIG.~\ref{fig:temporalmodes}\textbf{c} illustrates different strategies for utilizing multitone RAP pulses in the RAPPI protocol.\par

Experimentally, we investigate the spectral mode capacity of the RAPPI protocol by defining a set of distinct spectral modes/cells within the inhomogeneous broadening. FIG.~\ref{fig:temporalmodes}\textbf{d} illustrates the sequential storage of five spectral modes (top) and three spectral modes (bottom) using identical five-tone and three-tone RAP pulses, respectively. We observe that increasing the number of tones, leads to a reduced photon recall efficiency, likely due to off-resonant interactions of probe-pulse modes with RAP pulses addressing neighboring memory cells, or localized heating and decoherence arising from the elevated optical powers required for higher multitone RAP pulses.\par

Lastly, we encode different temporal modes across three distinct spectral memory cells (FIG.~\ref{fig:temporalmodes}\textbf{e}-\textbf{i}) and simultaneously store these spectro-temporal modes using a three-tone RAP pulse. Following a delay, a single-tone RAP pulse selectively recalls the temporal modes from only the chosen spectral cell (FIG.~\ref{fig:temporalmodes}\textbf{e}-\textbf{ii}). Crucially, a detailed experimental analysis shows no discernible cross-talk between the spectral memory cells (see Supplemental Material~\cite{Nasser2025Multimode}).

\begin{figure*}[htbp]
    \centering
    \includegraphics[width=0.87\linewidth]{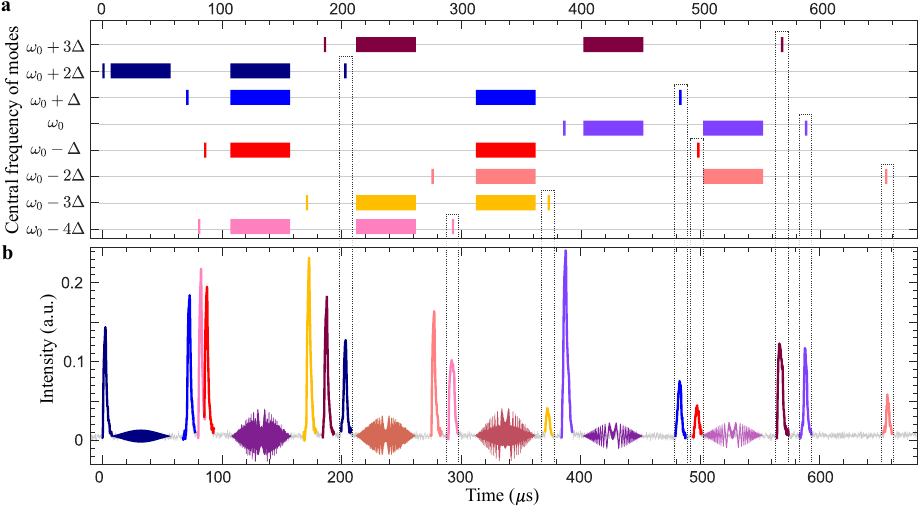}
   \caption{Optical random-access storage and retrieval of 8 spectral modes. \textbf{a,} Schematic representation of optical-RAM. After the memory cells are defined, input and output timings for each mode are randomly determined. Subsequently, the timings of the corresponding RAP pulses are calculated to execute random storage and on-demand recall of distinct spectral modes. Spectral memory cells are detuned by $\Delta = 3.5$~MHz, ensuring distinct frequency separation between modes. \textbf{b,} Experimental results for optical-RAM. Color-coded traces of input and output signals across different frequency bins, with dashed rectangles highlighting the output at each spectral bin. Each spectral input mode achieves a maximum peak power of 350 nW, with gated RAP pulses used to enhance visualization of the protocol.}
    \label{fig:8ModeORAM_and_Sequence}
\end{figure*}

Finally, we demonstrate optical-RAM utilizing combinations of single and multitone RAP pulses for write and read operations. Here, eight distinct spectral regions within the inhomogeneous broadening are designated as memory cells. The center frequency of the memory cells corresponds to the center frequency of the probe and RAP pulses for each mode, with the memory cells separated by $\Delta = 3.5$~MHz. FIG.~\ref{fig:8ModeORAM_and_Sequence} (\textbf{a}-schematic and \textbf{b}-experiment) illustrates the random-access storage and retrieval of spectral modes using different combinations of single- and multitone RAP pulses. Impressively, the RAPPI protocol facilitates the recall and storage of probe-pulses in any order and interspersed by RAP pulses addressing other spectral memory cells. By precisely timing RAP pulses and selecting corresponding spectral modes, we notably reduce the required number of write/read operations. In our case, only six sets of RAP pulses are required to implement optical-RAM with eight spectral modes. As a comparison, similar storage of eight photonic modes with EIT protocol or on-demand AFC QM typically requires about sixteen control pulses. Although the timing sequence (inputs/outputs and multitone RAP pulses) was determined manually, this process could be automated to support a larger number of modes. 

\section{Outlook}\label{sec:Conclusion_and_outlook}

We have presented a comprehensive theoretical and experimental demonstration of a novel method for multimode photonic qubit storage and recall in a Yb:YSO crystal. The RAPPI protocol introduced here offers a robust and straightforward approach that meets essential criteria for a QM in both quantum repeaters and random-access QM applications in quantum information processing systems.\par

Our findings emphasize the practicality of multitone RAP control pulses, which not only enhances the storage capacity, but also reduces the number of required write and read operations during on-demand photonic mode retrievals. The storage duration in this protocol is fundamentally limited by the optical $T_{2O}$ (considering only a two-level system and no dynamical decoupling or spin-wave storage), with optical depth also proving critical for overall protocol efficiency.\par

Both experimental and theoretical investigations indicate that FID decay restricts the signal-to-noise ratio for short storage times. This in turn puts a lower limit to the mean input photon number in the probe that can be discerned at the memory output. By overcoming the FID decay limits, we achieve successful storage of low photon number input pulses at extended durations. Moreover, similar to other on-demand QM protocols, we suggest that a slight spatial mismatch between input probe-pulse excitation and RAP pulses will enhance the SNR across all storage times.\par

Looking forward, this work sets the stage for further RAPPI protocol exploration in more complex quantum systems, including integration with emerging quantum technologies such as photonic quantum computing and advanced quantum networks. Future investigations may optimize the protocol across different material systems, assess higher-dimensional quantum states, and improve memory architecture for enhanced SNR, thus advancing the frontiers of QM applications in next-generation quantum information processing and communication implementations. \par

\begin{acknowledgments}
The authors would like to thank Khabat Heshami, Leili Esmaeilifar, and Farhad Rasekh for discussions on theoretical and experimental aspects of this research. This work was supported by the Government of Alberta Major Innovation Fund Project on Quantum Technologies, Alberta Innovates Advance Grant, Mitacs through the Mitacs Accelerate program, the Canadian Foundation for Innovation Infrastructure Fund (CFI-IF), the Natural Sciences and Engineering Research Council of Canada (NSERC) through the Alliance Quantum Consortia Grants CanQuest and ARAQNE.

\end{acknowledgments}

\appendix

\section{Experimental Details}

The Yb:YSO crystal employed in these experiments measures 5.5 $\times$ 6.5 $\times$ 12 mm, aligned with the crystallographic axes ${D_1} \times {D_2} \times {b}$. Light is incident along ${b}$ axis (polarization along ${D_2}$ axis) with both the front and rear surfaces not anti-reflection coated. A custom V-groove mount secures the crystal inside the cryostat, with alignment-free coupling of light into and out of the crystal achieved through 0.25-pitch GRIN lenses of 1 mm diameter. At the input, light from a single-mode (SM) 780~nm fiber is collimated by the first GRIN lens to a beam waist of about 200~$\mu$m. After crystal interaction, the light is collected into a SM 780~nm fiber at the output port utilizing an identical GRIN lens. The coupling efficiencies were measured as 35\% at room temperature and 15\% at 10~mK. \par

All experiments were conducted with the laser frequency set to the $|1_g\rangle  \rightarrow |4_e\rangle $ transition ($\omega_0$). Due to the repeated application of RAP pulses, during continuous execution of the memory protocol, the atomic population tends to equilibrate among the ground states. This, combined with the extended spin-lattice relaxation time, $T_{SLR}$, at the operational temperature of 10~mK, necessitates optical pumping (illustrated in Fig.~\ref{fig2:Setup}\textbf{b}) at regular intervals in order for experiments to ensure a consistent initial state. Specifically, the optical pumping initializes spin population predominantly in the $|1_g\rangle$ state prior to storing input probe-pulse, providing a controlled baseline for all measurements. The optical pumping sequence comprises a 1~GHz-wide chirped optical pulse targeting the $|3_g\rangle  \rightarrow |4 _e\rangle$ ($\omega_0 - 2\pi \times 2.37$~GHz) and $|4_g\rangle  \rightarrow |4_e\rangle $ ($\omega_0 - 2\pi \times 3.025$~GHz) transitions for 100~ms, followed by a single-frequency pulse applied to the $|2_g\rangle  \rightarrow |3_e\rangle $ ($\omega_0 - 2\pi \times 800$~MHz) transition for 50 ms. This cycle repeats until the complete optical pumping process reaches a duration of 300~ms. Frequency shifts from the central laser frequency to the designated optical pumping frequencies are produced by applying a serrodyne RF signal to a phase modulator (PM), achieving a single sideband frequency shift.\par

Following the optical pumping phase, the crystal undergoes a 6.5-second cooling period --- much shorter than the hour-long spin-relaxation time --- to dissipate residual internal heat before initiating the storage sequence. With a maximum storage duration of 700~$\mu$s, the full experimental cycle time per trial is approximately 7 seconds. The measured mixing-chamber plate (10~mK) temperature remains stable throughout; however, the cooling interval is crucial, as omitting this step results in decreased coherence times, indicating residual heat from the optical pumping that requires dissipation. \par

In the optical setup, two acousto-optic modulators (AOMs) are arranged in series to achieve an extinction ratio of approximately 66 dB between the $0^{\text{th}}$ and $1^{\text{st}}$ diffraction orders. For single-photon level experiments, additional neutral density filter are stacks in the beam path, providing an extinction ratio close to 100 dB. An Arbitrary Waveform Generator (AWG) supplies RF signals from two analog channels to the AOM and PM for the input, RAP, and optical pumping pulses. 

The single-tone and multitone RAP pulses are described by:
\begin{align}
    \begin{split}
    p(t)|_{f_{0}}= A_{0}\,\text{Sinc}\left(\frac{2t}{\tau_R}\right)\sin\left(2\pi f_{0}t + 2\pi\frac{\Delta_{R}}{\tau_R}t^{2}\right),\\
    \text{with } t \in \left[-\frac{\tau_R}{2}, \frac{\tau_R}{2}\right],
    \end{split}
\end{align}

\begin{align}
    \begin{split}
     P(t)|_{f(\Delta)}= p(t)|_{f_{0}} + p(t)|_{f_{0}-\Delta} +  p(t)|_{f_{0}+\Delta} + \dots,\\
     \text{with } t \in \left[-\frac{\tau_R}{2}, \frac{\tau_R}{2}\right],
    \end{split}
\end{align}
where $A_0$ is the maximum pulse amplitude, $\tau_R = 50$ µs, $f_0$ represents the driving frequency of the AOM (200~MHz), $\Delta_{R}$ is the chirp range, and $\Delta$ denotes the detuning defined for memory cells, set to $\Delta = 3.5$~MHz in this study.

Two outputs from an Arbitrary Function Generator (AFG) are routed through identical electronic switches, gated by AWG-generated digital markers, to drive the first and third AOMs. As described in the main text, the third AOM functions as a gate for the detector, preventing detection of the RAP pulses. Though these pulses are lower in power compared to conventional short, high-power $\pi$-pulses, they remain capable of saturating the NewPort 2053 photodetector used in this study. This detector, with a peak gain of $18.8 \times 10^6$ V/W, can detect pulses with duration of $<$500 ns and peak powers as low as 1-10 nW.

Temporal modes are generated by square pulses with a specified duty cycle (1~$\mu$s on, 2~$\mu$s off in this case). For simultaneous, randomly encoded spectro-temporal modes, a custom code initially prepares six temporal modes for each spectral component ($\omega_0$ and $\omega \pm \Delta$). The code then removes up to a certain number of pulses from each spectral mode randomly, making them distinguishable from one another and outputs a simultaneous spectro-temporal modes.

A unique attribute of the 980~nm laser (Toptica DLpro, linewidth of 100 KHz) used in this study is its stability during storage experiments, eliminating the need for active frequency locking. This property benefits the RAPPI protocol, distinguishing it from other QM protocols that require high laser stability. During minor lab temperature fluctuations impacting laser stability, a shift of approximately $\pm~1–2~\mu$s in echo timing was observed. To confirm that this shift originated from laser frequency fluctuations, we locked the laser to a high-finesse external reference cavity (Stable Laser System) 600~MHz detuned from the atomic resonance and verified that echo timing returned precisely to the expected values.

To assess the photon-number of the input probe-pulse in the single photon regime, we carefully account for all the propagation losses in our setup (FIG.~\ref{fig2:Setup}). Specifically, we account for the 15\% efficiency of coupling in and out of the Yb:YSO crystal to the optical fiber, the 60\% diffraction efficiency of the gating AOM, and subsequent 65\% coupling efficiency back into an optical fiber, and finally the 12\% detection efficiency of the APD at 980~nm. All combined we calibrate that the detection of one photon with the APD correlates to approximately 150 input photons. Based on the measured count rates on the APD, when the optical pulse is off-resonant with the atoms, we assess that 2500 photons reach the crystal, resulting in the detection of 16-20 photons per pulse.\par

Data analysis is performed using two methods. For the measurements shown in FIGs.~\ref{fig:efficiency}, ~\ref{fig:temporalmodes}, and~\ref{fig:8ModeORAM_and_Sequence}, for each storage time setting we recorded the amplitudes of twenty trials, and select the trial yielding the maximum echo amplitude for analysis. In contrast, for single-photon level pulse storage (as shown in FIG.~\ref{fig:Single_photon_level}), analysis involved averaging across 50 consecutive recorded histograms.

\bibliography{References}

\end{document}